\documentclass[aps,pre,superscriptaddress,reprint]{revtex4-1}
\usepackage[utf8]{inputenc}
\usepackage[T1]{fontenc}
\usepackage[english]{babel}
\usepackage{graphicx} 
\usepackage{xcolor}
\usepackage{geometry}
\usepackage{amsmath,amssymb}
\usepackage{diagbox}

\newcommand{\toff}{\ensuremath{T_{\text{off}}}}
\newcommand{\Eb}{\ensuremath{E_\text{b}}}
\newcommand{\vect}[1]{\ensuremath{\mathbf{#1}}}
\newcommand{\dist}{\ensuremath{a}}
\newcommand{\parody}{\textsc{parody}}
\newcommand{\ri}{\ensuremath{r_\text{i}}}
\newcommand{\ro}{\ensuremath{r_\text{o}}}
\newcommand{\rot}{\ensuremath{\Omega}}
\newcommand{\drot}{\ensuremath{\Delta\Omega}}
\newcommand{\reynolds}{\ensuremath{\mathrm{Re}}}
\newcommand{\prandtl}{\ensuremath{\mathrm{P_m}}}
\newcommand{\eckman}{\ensuremath{\mathrm{E}}}
\newcommand{\ratio}{\ensuremath{\chi}}
\newcommand{\reymag}{\ensuremath{\mathrm{Rm}}}
\newcommand{\rmc}{\ensuremath{\mathrm{Rm}_\text{c}}}
\newcommand{\ari}{\ensuremath{\theta}}
\newcommand{\muzero}{\ensuremath{\mathrm{\mu}_0}}
\newcommand{\laplacien}{\ensuremath{\vect{\nabla^2}}}
\newcommand{\ez}{\ensuremath{\vect{e}_\text{z}}}

\begin{document}

\renewcommand{\figurename}{FIG.}
\renewcommand{\tablename}{TABLE.}

\title{Intermittency in spherical Couette dynamos}

\author{Rapha\"el~Raynaud}
\email{raphael.raynaud@ens.fr}
\affiliation{MAG (LRA), \'{E}cole normale supérieure, 24 rue Lhomond, 75252
  Paris Cedex 05, France}
\affiliation{LERMA, CNRS UMR 8112}

\author{Emmanuel~Dormy}
\email{dormy@phys.ens.fr}
\affiliation{MAG (LRA), \'{E}cole normale supérieure, 24 rue Lhomond, 75252
  Paris Cedex 05, France}
\affiliation{IPGP, CNRS UMR 7154}

\date{Received 18 December 2012; revised manuscript received 22
  February 2013; published 18 March 2013}

\begin{abstract}
We investigate dynamo action in three-dimensional numerical
simulations of turbulent spherical Couette flows. Close to the onset
of dynamo action, the magnetic field exhibits an intermittent
behavior, characterized by a series of short bursts of the magnetic
energy separated by low-energy phases. We show that this behavior
corresponds to the so-called on-off intermittency. This behavior is
here reported for dynamo action with realistic boundary conditions. We
investigate the role of magnetic boundary conditions in this
phenomenon.
\end{abstract}


\maketitle

\section{Introduction}

First suggested by Joseph Larmor in 1919, dynamo action, i.e. the
self-amplification of a magnetic field by the flow of an electrically
conducting fluid, is considered to be the main mechanism for the
generation of magnetic fields in the universe for a variety of
systems, including planets, stars, and galaxies~\cite{book_dormy}.
Dynamo action is an instability by which a conducting fluid transfers
part of its kinetic energy to magnetic energy.

In experiments, it is rather difficult to achieve a regime of
self-excited dynamo action. The low value of the magnetic Prandtl
number of liquid metals requires the injection of a sufficiently high
mechanical power, and thus generates turbulent flows, before reaching
the dynamo threshold. Dynamo action was first observed experimentally
only in 2001, in Karlsruhe~\cite{karlsruhe} and Riga~\cite{riga}, and
then in 2007 with a von K\'arm\'an swirling flow of liquid
sodium~\cite{berhanu}.

In parallel with these approaches, numerical simulations have been
carried out to model either laboratory experiments or astrophysical
systems, for which the spherical geometry is relevant. We investigate
spherical Couette flow and focus on the characteristics of the magnetic
field close to the dynamo onset. We observe a series of short bursts of
the magnetic energy separated by low-energy phases. This intermittent
behavior, also known as on-off intermittency or blowout bifurcation,
is usually interpreted as the effect of a multiplicative noise acting
on a bifurcating system~\cite{fujisaka86,platt}.

On-off intermittency has so far never been observed in dynamo
experiments, except in the case of an externally amplified magnetic
field~\cite{verhille}. In contrast, it has been reported in a small
number of numerical simulations~\cite{sweet1,leprovost,alex}, all
relying on a flow in a periodic geometry produced by a periodic
analytic forcing. Here we investigate the influence of a realistic
choice of boundary conditions on this phenomenon.

\section{Governing equations}\label{s_eq}

The spherical Couette flow geometry consists of two concentric spheres
in differential rotation: the outer sphere, of radius~\ro{}, is
rotating around the vertical axis~\ez{} with an angular
velocity~\rot{}, and the solid inner sphere, of radius~\ri{}, is
rotating at velocity $\rot+\drot$ around an axis that can make an
angle~\ari{} with~\ez{}.  The aspect ratio $\ratio = \ri/\ro$ is set
to 0.35 to mimic that of Earth's liquid core. The spherical shell in
between the two spheres is filled with an incompressible conducting
fluid of kinematic viscosity~$\nu$, electrical conductivity~$\sigma$,
and density~$\rho$. Its magnetic permeability~\muzero{} is that of
vacuum. The magnetic diffusivity~$\eta$ is defined as~$\eta =
1/(\muzero\sigma)$.

We describe the problem in the reference frame rotating with the outer
sphere.  This introduces two extra terms in the governing equations:
the Coriolis force and the centrifugal acceleration. The latter can be
rewritten in the form $\tfrac{1}{2}\nabla \left(\Omega ^2 s^2\right)$,
where $s$ denotes the distance to the axis of rotation. This term is a
gradient and can be added to the pressure term which acts as a
Lagrange multiplier to enforce the solenoidal condition on the
velocity field.  To establish the set of equations for this system, we
rely on the same non-dimensional form as in~\cite{celine}: the
velocity~\vect{u} is scaled by $\ri\drot$, the magnetic field \vect{B}
by $\sqrt{\rho\muzero\ri\ro\left(\rot + \drot\right)\drot}$, and the
length scale by \ro{}. The Navier-Stokes equation governing the fluid
velocity~\vect{u} then takes the form
\begin{multline}\label{eq_system_ns}
  \frac{\partial\, \vect{u}}{\partial t} + \left( \vect{u}\cdot
  \nabla\right)\vect{u} + \frac{2}{\eckman\reynolds} \left(
  \ez\times\vect{u}\right) = -\frac{1}{\reynolds}\nabla\Pi + \\
  \frac{1}{\reynolds}\laplacien\vect{u} + \frac{1}{\reynolds}\left(
  \frac{1}{\eckman}+\frac{\reynolds}{\ratio}\right)\left(
  \nabla\times\vect{B}\right) \times\vect{B} \,,
\end{multline}
and the induction equation for the magnetic field \vect{B}, 
\begin{equation}\label{eq_system_in}
  \frac{\partial\,
    \vect{B}}{\partial t} = \nabla\times\left(
  \vect{u}\times\vect{B}\right) + \frac{1}{\reymag}\laplacien\vect{B}
  \,.
\end{equation}
%%%
Both fields are solenoidal
\begin{equation}\label{eq_solenoid}
\nabla\cdot\vect{u}=0 \, , \qquad
\nabla\cdot\vect{B}=0 \, .
\end{equation} 
%%%

The dimensionless parameters are the Ekman number $\eckman =
\nu/\left( \rot\ro^2\right)$, the Reynolds number $\reynolds = \left(
\ro\ri\drot\right) /\nu$, the magnetic Prandtl number $\prandtl =
\nu/\eta$, and the magnetic Reynolds number $\reymag =
\reynolds\prandtl$. The potential~$\Pi$ includes all gradient terms
(the pressure term as well as the centrifugal effect introduced
above).  The Reynolds number varies with the rotation rate of the
inner sphere, while the Ekman number is inversely proportional to the
rotation rate of the outer sphere.  When the latter is at rest, the
Ekman number tends toward infinity and the Coriolis term in the
Navier-Stokes equation vanishes. In our simulations, the Ekman number
is set to $10^{-3}$. This moderate value yields a moderate computing
time.

We impose no slip boundary conditions for the velocity field on both
spheres.  Magnetic boundary conditions are of three types.  The first
one can only be applied to the inner sphere, as it implies a meshing
of the bounding solid domain. The inner sphere can be a conductor with
the same electric and magnetic properties as the fluid. In that case
the magnetic diffusion equation is discretized and solved in the solid
conductor (we refer to this set of boundary conditions as
``conducting'').  The outer sphere as well as the inner sphere can be
electrical insulators. In that case the magnetic field is continuous
across the boundary and matches a potential field, decaying away from
the boundary. The spherical harmonic expansion allows an explicit and
local expression for these boundary conditions (we refer to these
boundary conditions as ``insulating'').  In addition, the use of
high-magnetic-permeability boundary conditions may enhance dynamo
action~\cite{vks_ferro_effects}. Therefore, we also used boundary
conditions which enforce the magnetic field to be normal to the
boundary. This is equivalent to assuming that the medium on the other
side of the boundary has an infinitely larger permeability (we refer
to these boundary conditions as ``ferromagnetic'').  The different
configurations investigated in this study are summarized in
Table~\ref{t_bc}.

\begin{table}[htbp]
  \begin{center}
    \begin{tabular}{l c c}
      \hline\hline
              & Inner Sphere  & Outer Sphere \\
      \hline
      B.C.1 & Conducting    & Insulating \\ 

      B.C.2 & Insulating    & Insulating \\

      B.C.3 & Ferromagnetic &  Ferromagnetic \\
      \hline\hline
    \end{tabular}
  \caption{The three different configurations of 
    magnetic boundary conditions used in this study.}\label{t_bc}
  \end{center}
\end{table}

We integrated our system with \parody{}~\cite{parody}, a parallel code
which has been benchmarked against other international codes.  The
vector fields are transformed into scalars using the poloidal-toroidal
decomposition.  This expansion on a solenoidal basis enforces the
constraints~\eqref{eq_solenoid}.  The equations are then discretized
in the radial direction with a finite-difference scheme on a stretched
grid.  On each concentric sphere, variables are expanded using a
spherical harmonic basis (i.e., generalized Legendre polynomials in
latitude and a Fourier basis in longitude). The coefficients of the
expansion are identified with their degree $l$ and order $m$.  The
simulations were performed using from 150 to 216 points in the radial
direction, and the spherical harmonic decomposition is truncated at
$(l_\text{max},m_\text{max}) = (70,20)$. We observe for both spectra a
decrease of more than two orders of magnitude over the range of $l$
and $m$. This provides an empirical validation of convergence.  We
checked on a few critical cases that the results are not affected when
the resolution is increased to $l_{\text{max}}=100\,$.

Let us define the non-dimensional kinetic and magnetic energy
densities as
\begin{equation}
E_k=\frac{1}{V_s} \int_{V_s} \vect{u}^2 {\rm d} \vect{x} \, ,
\end{equation} 
\begin{equation}
E_b=\frac{1}{V_s} \, \frac{1}{\reynolds}\left(
\frac{1}{\eckman}+\frac{\reynolds}{\ratio}\right)\, \int_{V_s}
\vect{B}^2 {\rm d} \vect{x} \, ,
\end{equation} 
in which the unit of energy density is $\rho \left( r_i \, \Delta
\Omega \right)^2 $. In the above expressions, $V_s$ refers to the
volume of the spherical shell. In addition, we also investigate the
symmetry of the flow and the symmetry of the magnetic field with
respect to the equatorial plane.  To that end, we define the
contributions to the energy densities corresponding to the symmetric
and antisymmetric components of the velocity (respectively $E_{kS}$
and $E_{kA}$) and magnetic field (respectively $E_{bS}$ and $E_{bA}$).
The symmetric and antisymmetric contributions to the kinetic energy
density respectively correspond to the flows
\begin{equation}
\vect{u}_S(x,y,z) = \tfrac{1}{2} \left[ \vect{u}(x,y,z) +
\vect{u}(x,y,-z) \right],
\end{equation}
\begin{equation}
\vect{u}_A(x,y,z) = \tfrac{1}{2} \left[ \vect{u}(x,y,z) -
\vect{u}(x,y,-z) \right].
\end{equation} 
In contrast, the symmetries are reversed for the magnetic field. This
comes from the fact that the magnetic field is a pseudovector
(i.e, the curl of a vector). Then,
\begin{equation}
\vect{B}_S(x,y,z) = \tfrac{1}{2} \left[ \vect{B}(x,y,z) -
\vect{B}(x,y,-z) \right] ,
\end{equation} 
\begin{equation}
\vect{B}_A(x,y,z) = \tfrac{1}{2} \left[ \vect{B}(x,y,z) +
\vect{B}(x,y,-z) \right] .
\end{equation} 
According to our definition, the dipolar component is symmetric.

\section{Direct numerical simulations}\label{s_results}

As shown by~\cite{celine}, contra-rotation is more efficient than
co-rotation for dynamo action. In order to introduce more control over
the system, we let the angle~\ari{} between the axes of rotation of
both spheres take any value in $\left[ 0, \pi \right]$.  Contrary to
our expectations, we do not significantly lower the dynamo threshold
with the inclination of the rotation axis of the inner sphere. In
fact, for $\ari=\pi/2$, the fluid is mainly in co-rotation with the
outer sphere, and dragged only by a thin layer on the inner sphere,
which is not sufficient to trigger dynamo action.  In our parameter
regime, the best configuration seems to remain $\ari=\pi$, when the
two spheres are in contra-rotation. We therefore keep this parameter
fixed in the rest of the study.

\subsection{Role of boundary conditions}

\begin{figure*}[p]
  \begin{center}
    \includegraphics[width=0.4\textwidth, clip=true, trim=0 180 0
      180]{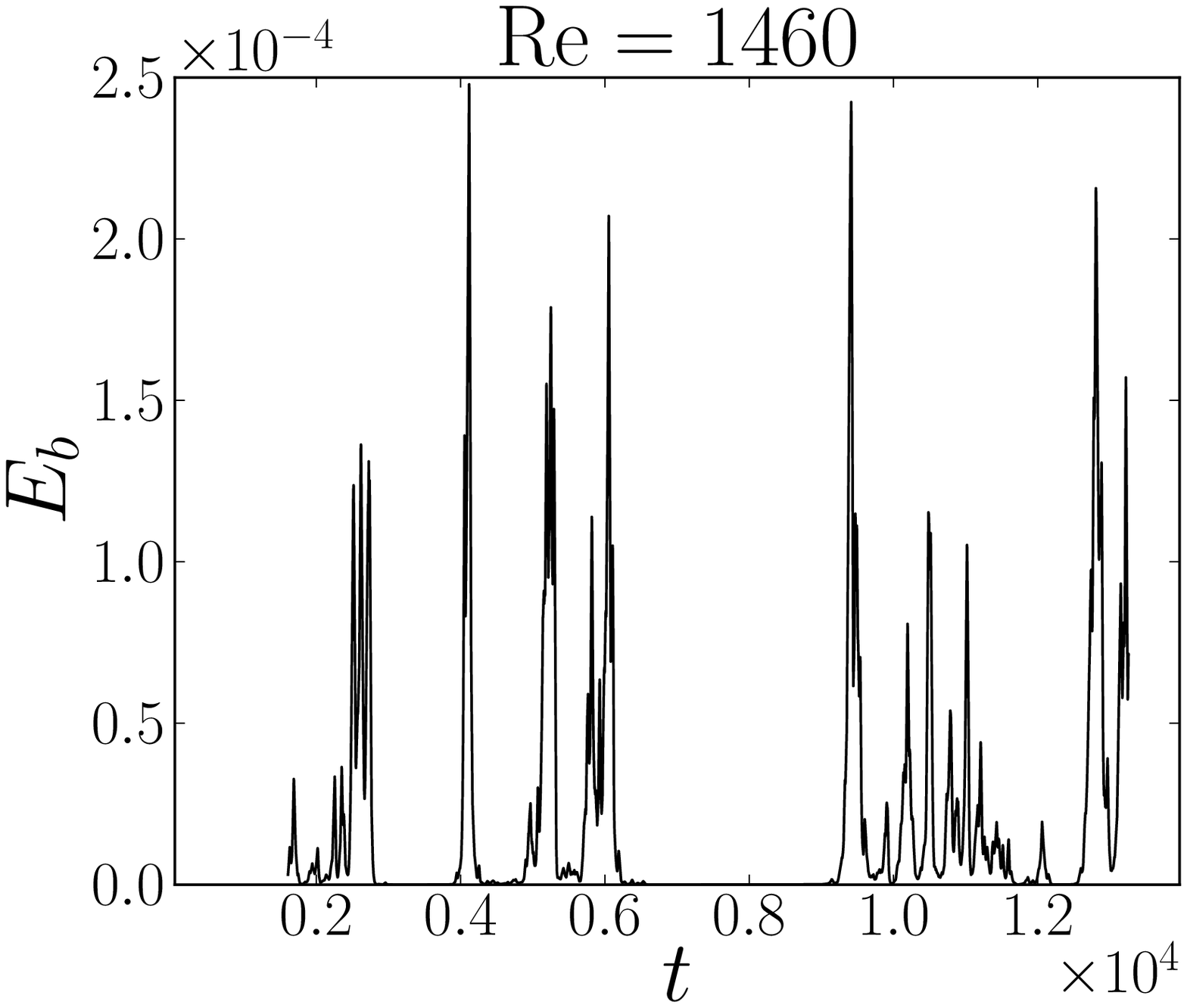}%
    \includegraphics[width=0.4\textwidth, clip=true, trim=0 180 0
      180]{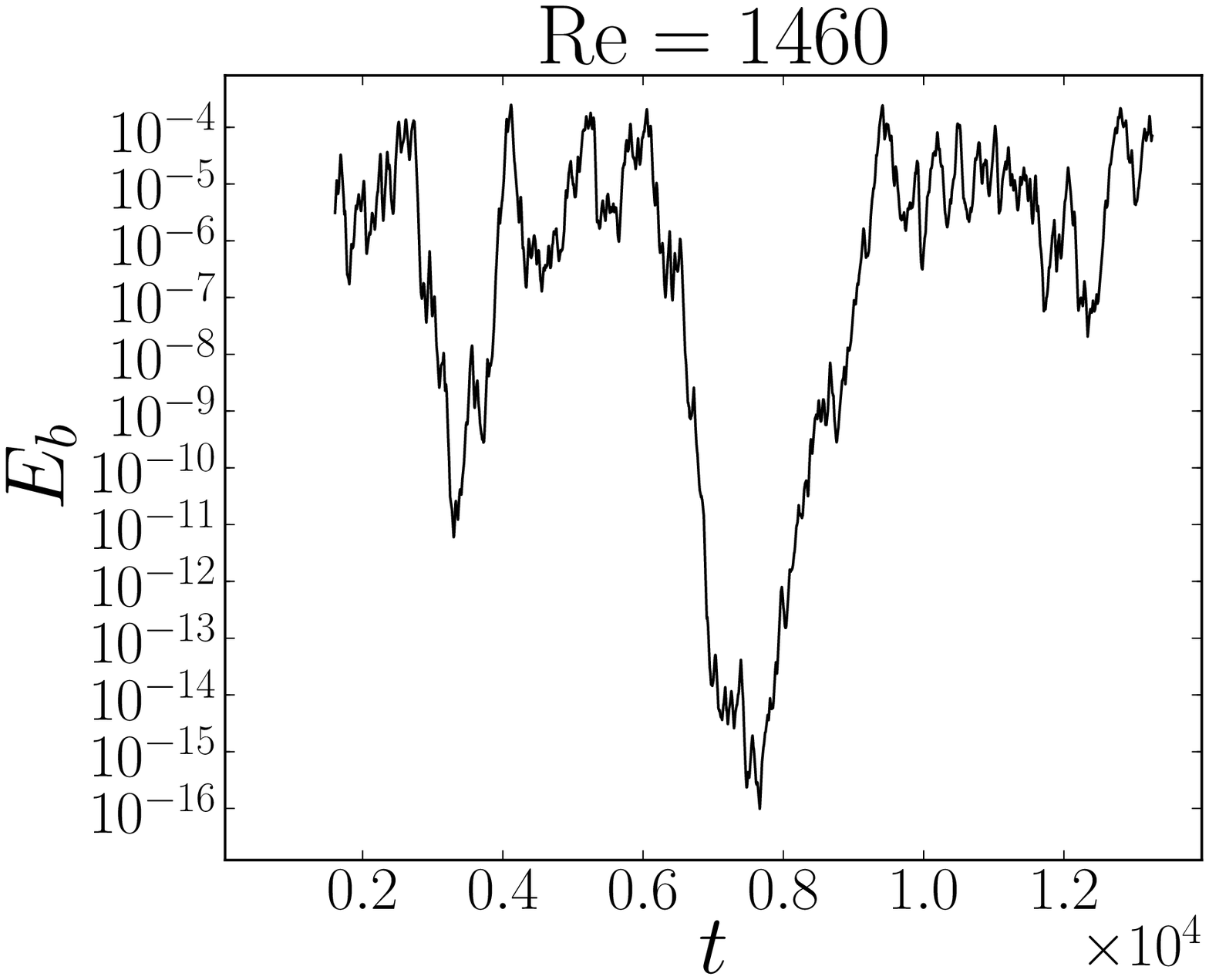}%

    \includegraphics[width=0.4\textwidth, clip=true, trim=0 180 0
      180]{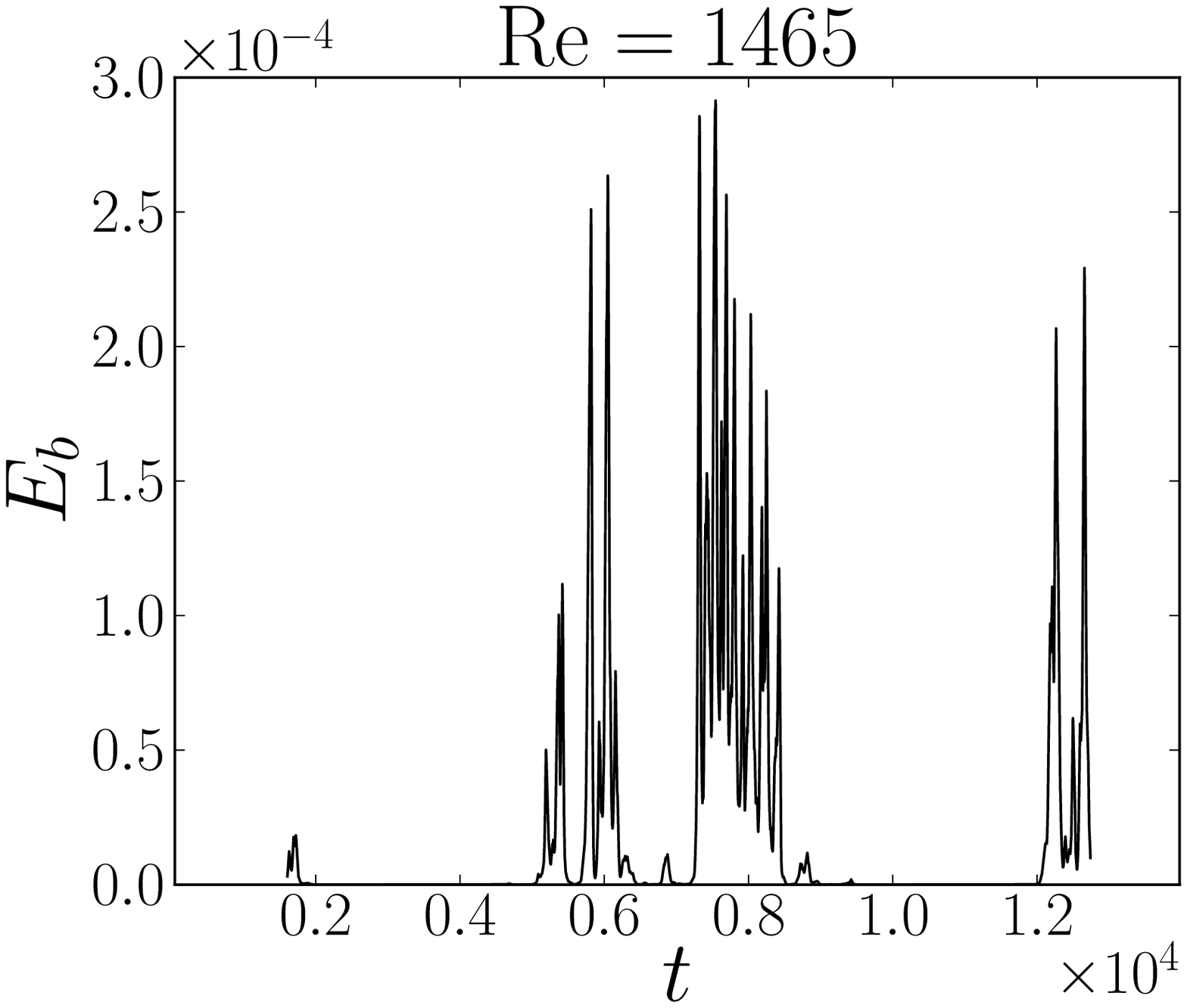}%
    \includegraphics[width=0.4\textwidth, clip=true, trim=0 180 0 
      180]{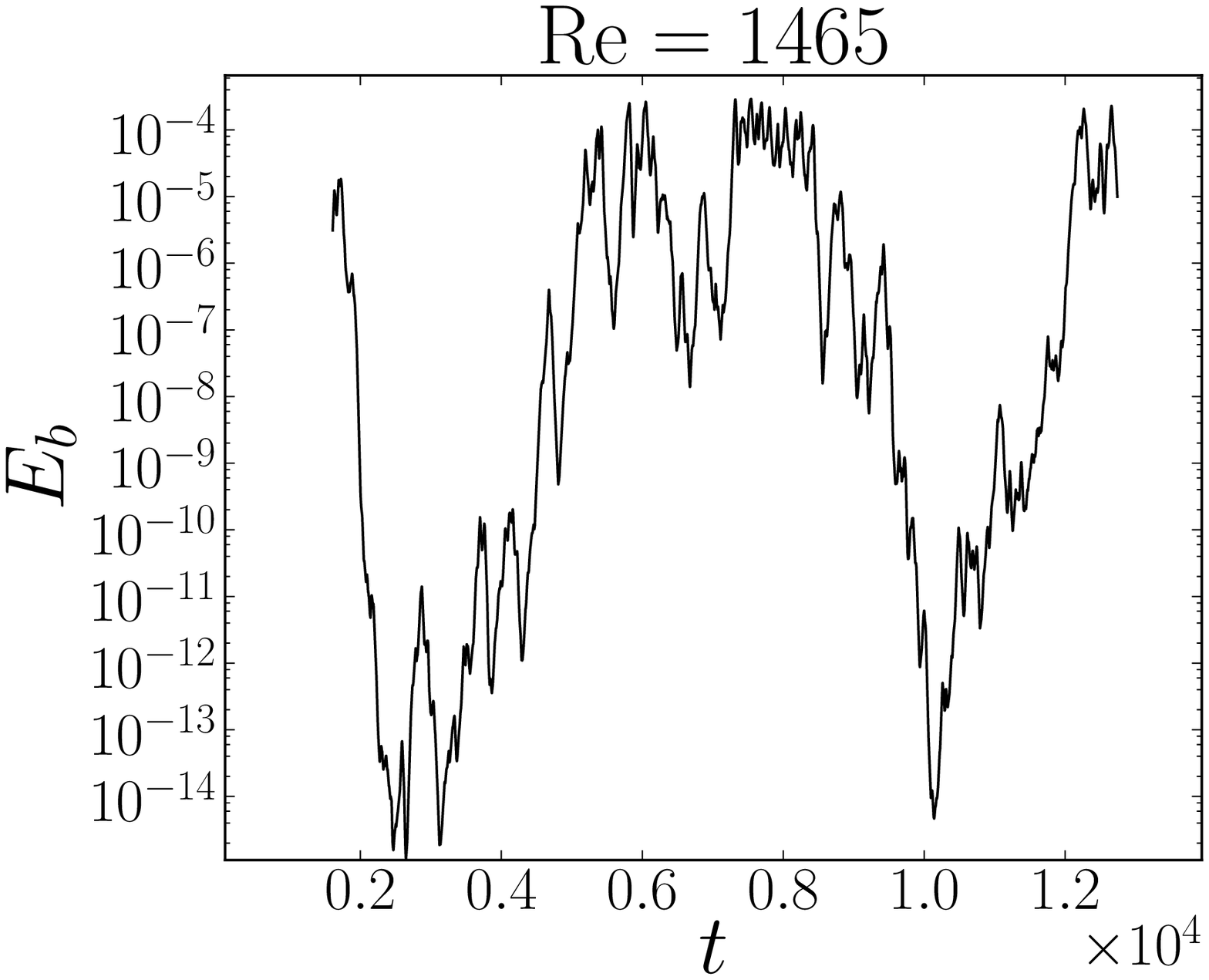}%

    \includegraphics[width=0.4\textwidth, clip=true, trim=0 180 0
      180]{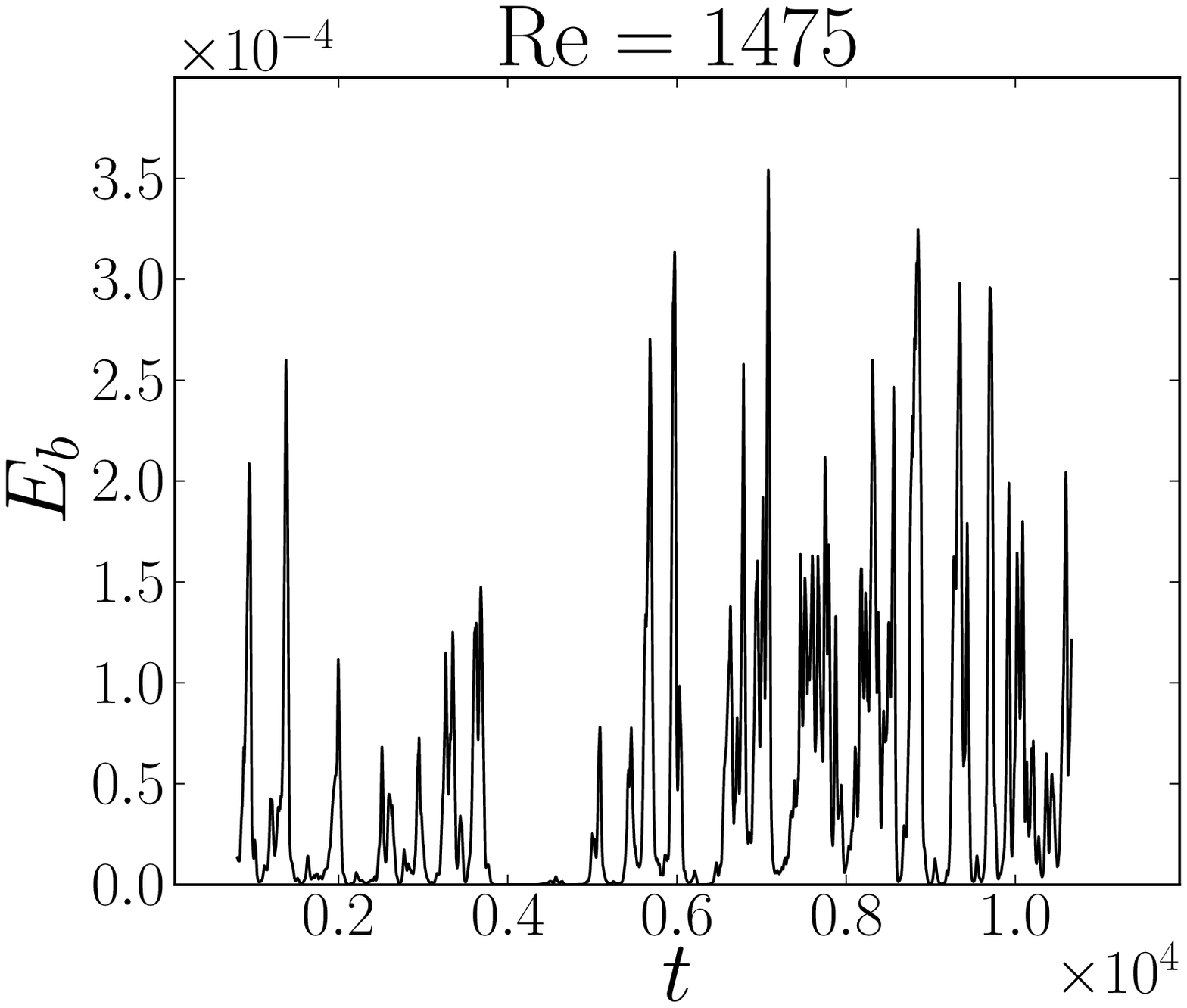}%                                           
    \includegraphics[width=0.4\textwidth, clip=true, trim=0 180 0
      180]{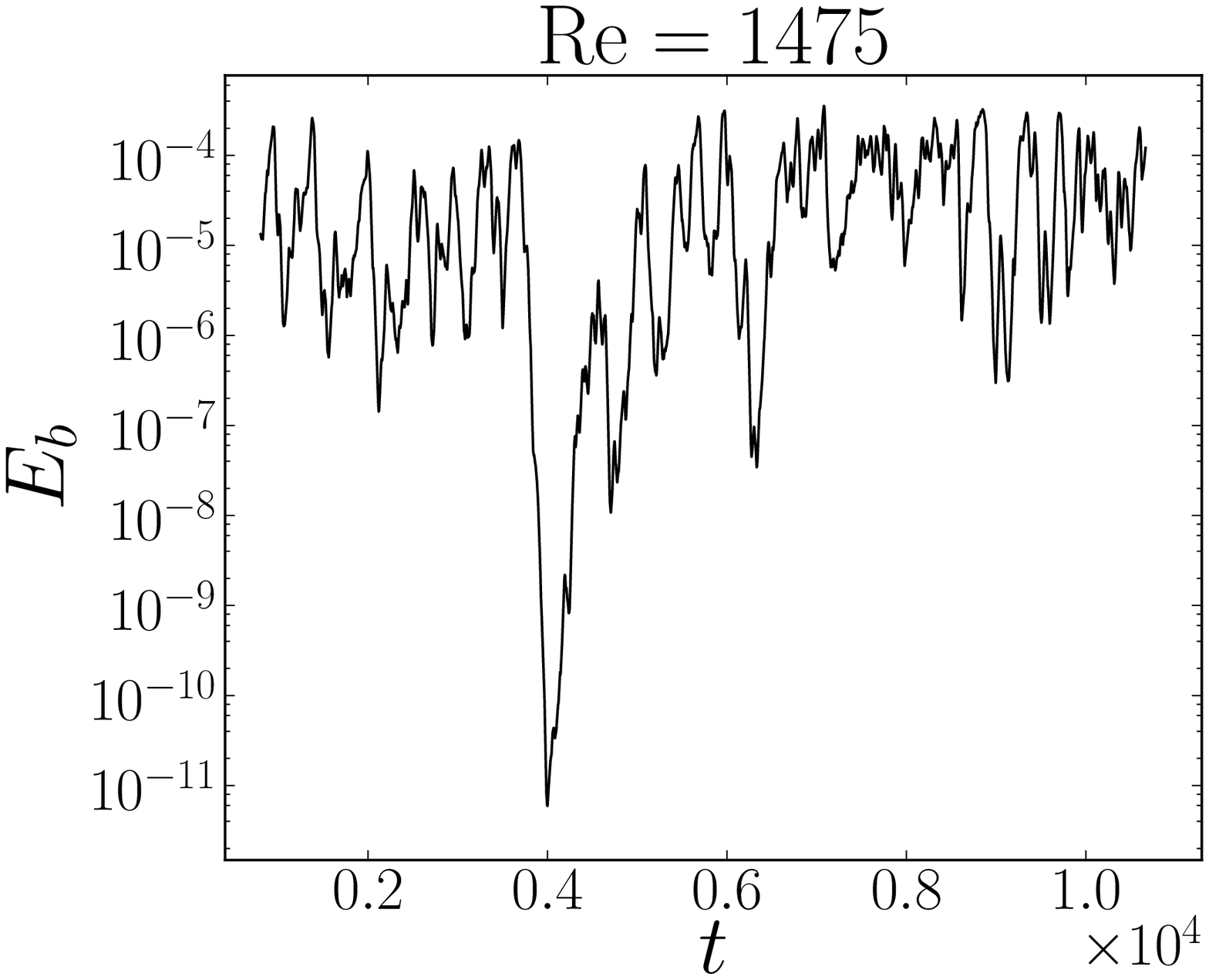}%                                           
                                                              
    \includegraphics[width=0.4\textwidth, clip=true, trim=0 180 0
      180]{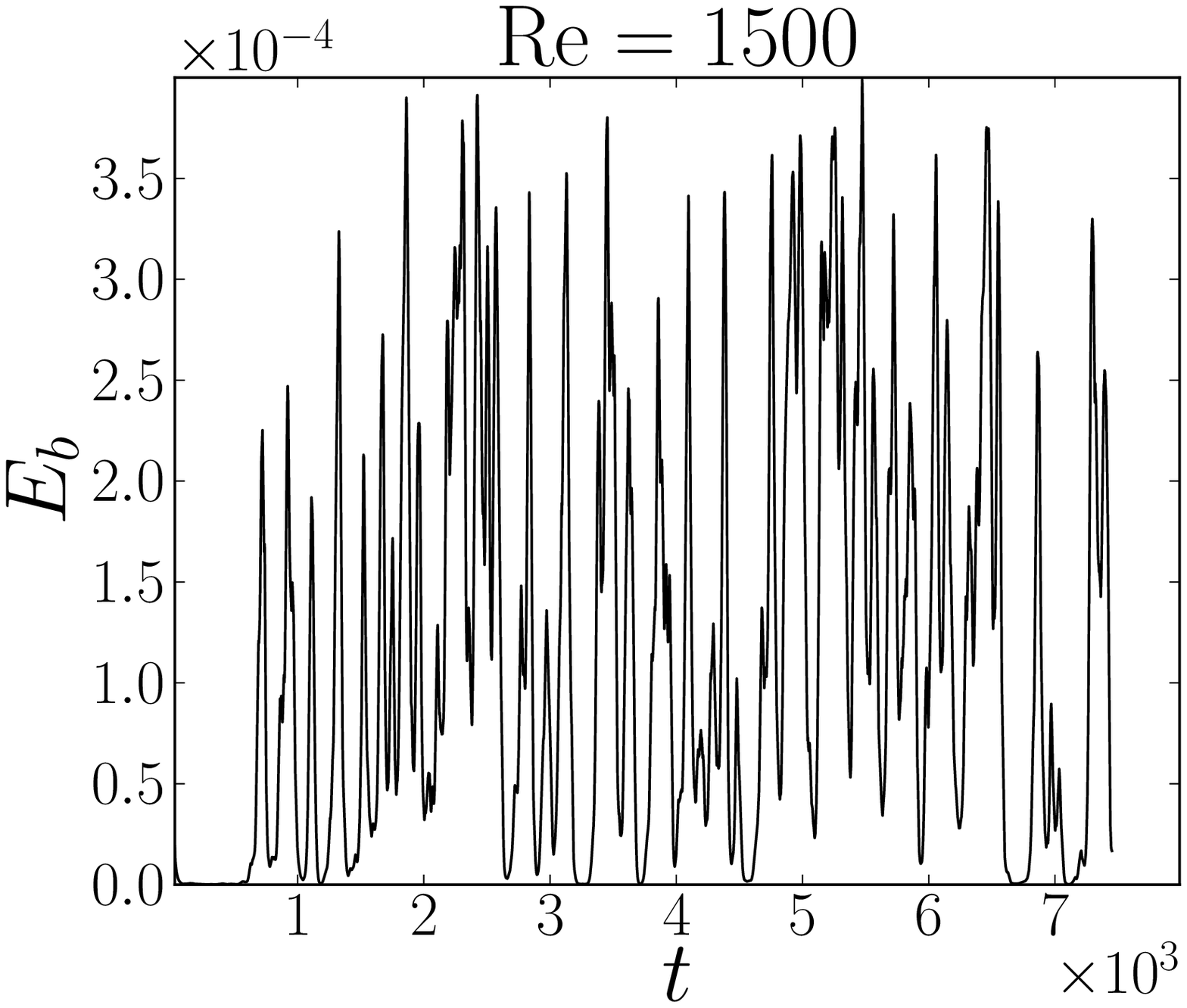}%                                          
    \includegraphics[width=0.4\textwidth, clip=true, trim=0 180 0
      180]{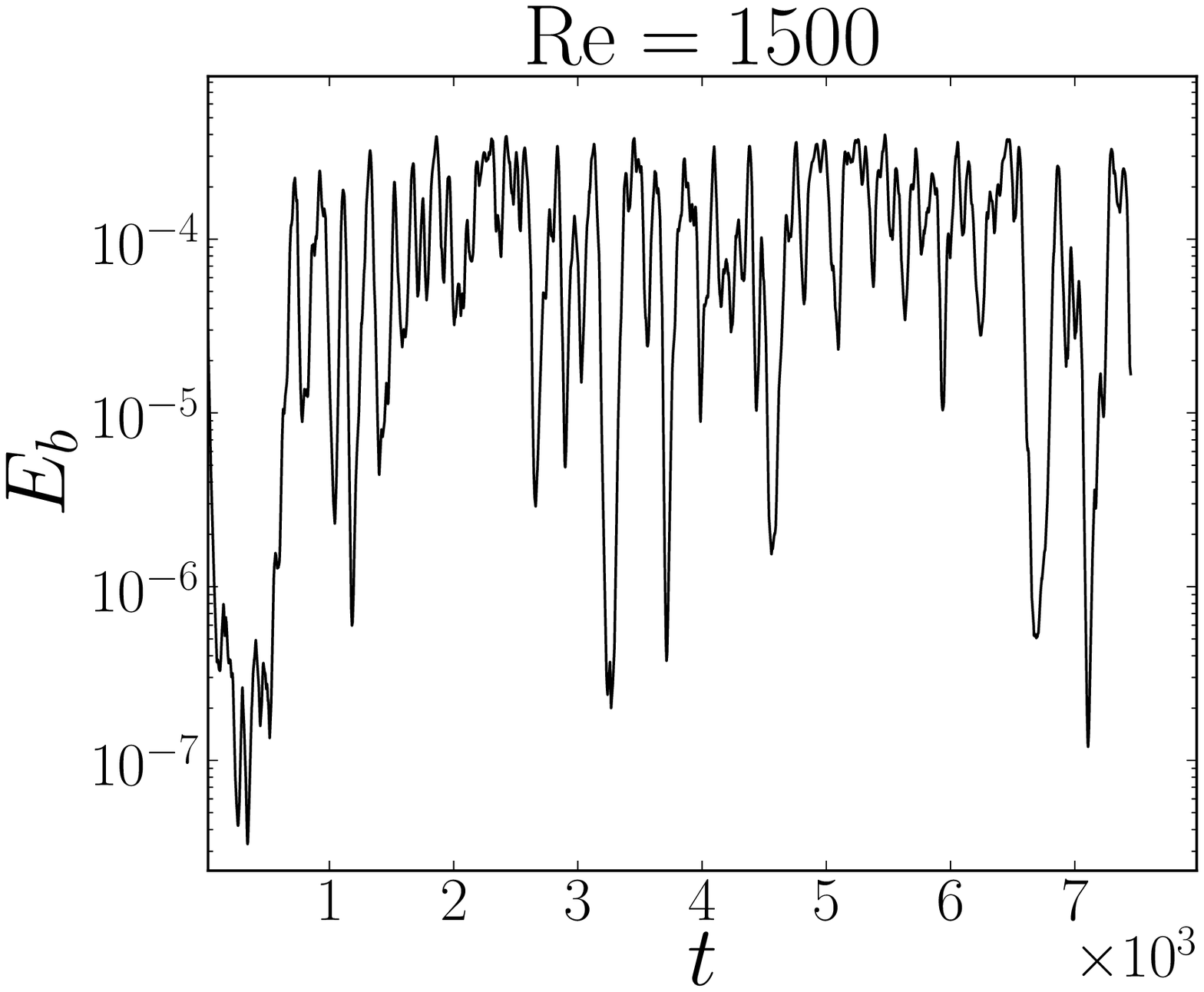}%

    \caption{Time evolution of the magnetic energy in linear (left)
      and log scale (right) for increasing Reynolds numbers at
      $\prandtl=0.2$, using a conducting inner sphere (B.C.1 in
      Table~\ref{t_bc}). At lower Reynolds numbers (top), we see in
      linear scale the characteristic intermittent bursting.
      Intermittency gradually disappears at higher Reynolds numbers
      and the field reaches saturation.}\label{f_onoff}
  \end{center}
\end{figure*}

Let us first investigate the dynamo transition in this setup at fixed
magnetic Prandtl number $\prandtl=0.2$, using the Reynolds number as
the controlling parameter. With a conducting inner sphere and an
insulating outer sphere, we find a critical magnetic Reynolds number $
\rmc \in [291.0; 292.0] $, which is in good agreement
with~\cite{celine}. Above the dynamo onset, the magnetic field
displays an intermittent behavior characterized by series of short
bursts of the magnetic energy separated by low energy phases (see
Fig.~\ref{f_onoff}). When the distance to the threshold increases,
bursts become more and more frequent and eventually intermittency
disappears.

Changing the boundary conditions generally leads to different
thresholds for dynamo action.  Using ferromagnetic boundary
conditions, we find a critical magnetic Reynolds number
$\rmc \in [298.6; 300.0]$.  With insulating boundary conditions, the
threshold becomes large and involves larger numerical resolutions. In
order to maintain the hydrodynamic Reynolds number at values which
involve a moderate resolution, we therefore had to increase the
magnetic Prandtl number from 0.2 to $0.4\,$.  We then obtain the
dynamo onset for $\rmc \in [530.0; 534.8] \,$.  We emphasize that we
observe the same intermittent regime with all the above choices of
boundary conditions as long as the magnetic Reynolds number is close
enough to the onset of the instability.

For all boundary conditions, we observe that the dominant mode is
predominantly of quadrupolar symmetry [the larger poloidal and
toroidal modes are the $(l=2, m=0)$ and $(l=1, m=0)$ modes,
respectively]. For these Reynolds numbers, the flow is predominantly
equatorially symmetric ($E_{kA} \ll E_{kS}$).

\subsection{Increasing the magnetic Prandtl number}

Having assessed that the intermittent behavior of the magnetic field
near onset could be observed with three different sets of boundary
conditions, we restrict here our attention to simulations with
ferromagnetic boundary conditions.

\begin{figure}[htbp]
  \centering
  \includegraphics[width=0.4\textwidth,clip=true, trim=0 180 0
    180]{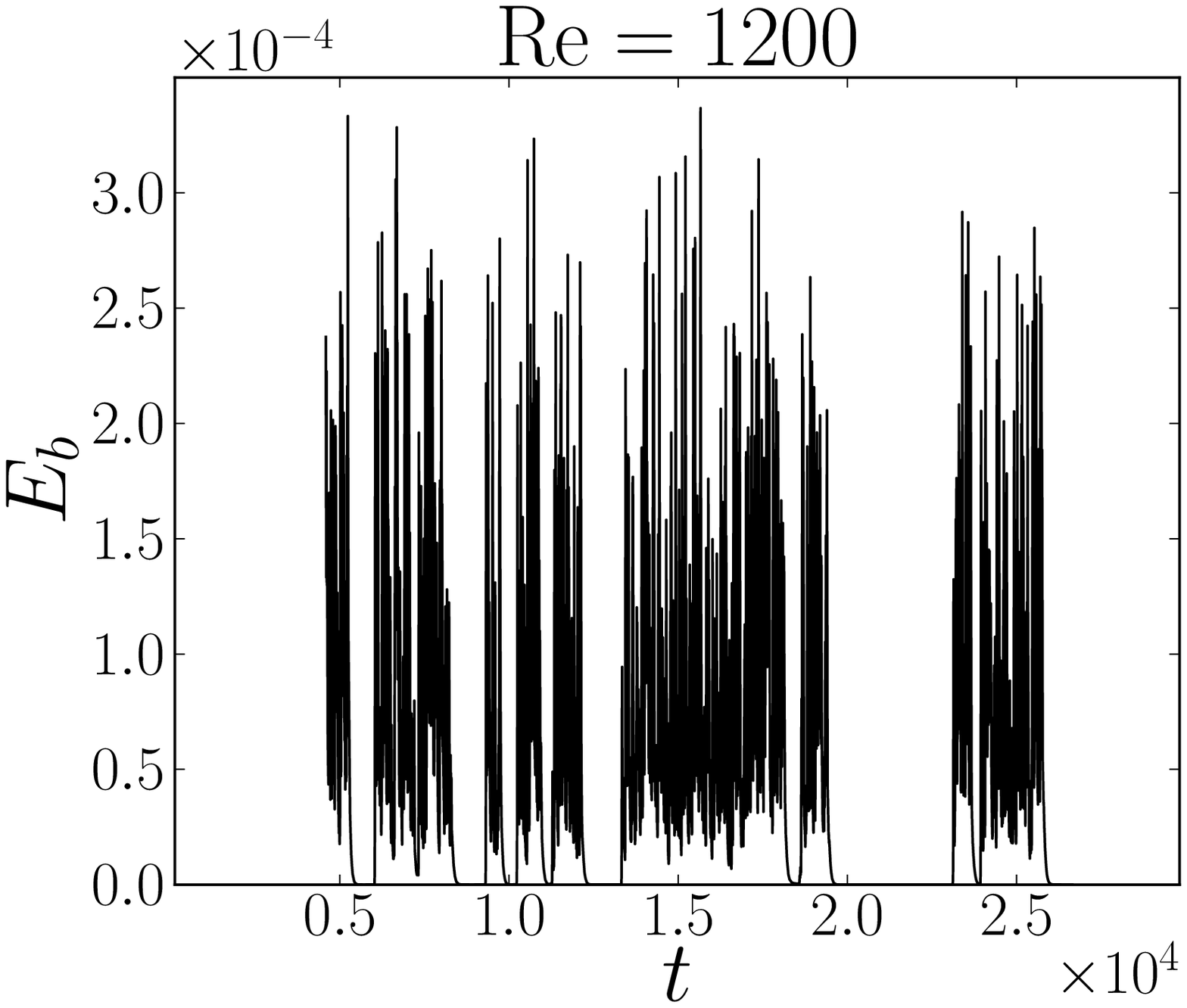}
  \includegraphics[width=0.4\textwidth,clip=true, trim=0 180 0
    180]{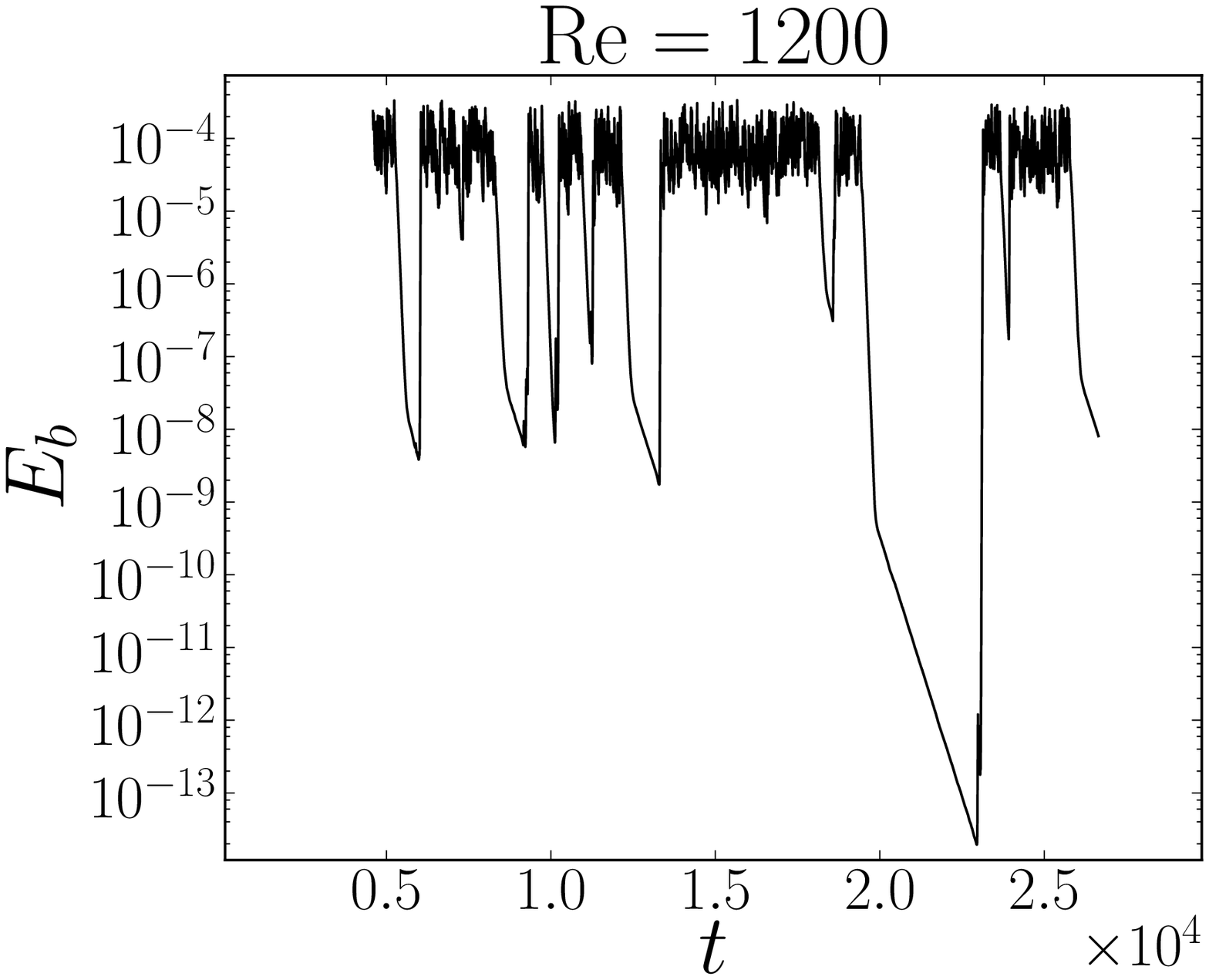}
  \caption{Time evolution of the magnetic energy in
    linear and log scales for $\prandtl=2$ and $\reynolds=1200$, using
    ferromagnetic boundary conditions. Instead of bursts, we now
    observe phases of dynamo activity which seem to randomly alternate
    with phases of exponential decay. The latter are no longer chaotic
    and are instead characterized by two different decay
    rates.}\label{f_hautpm}
\end{figure}

\begin{figure}[htbp]
  \centering
  \includegraphics[width=0.4\textwidth,clip=true]{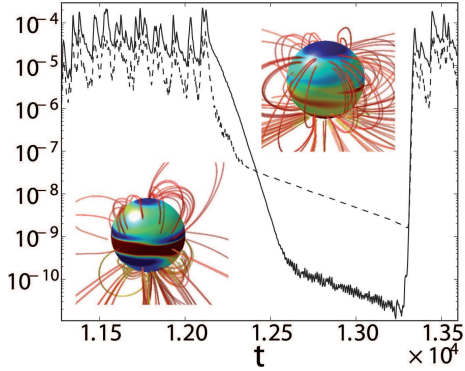}
  \caption{(Color online) Time evolution of the symmetric (dashed
    line) and antisymmetric (solid line) part of the magnetic
    energy. We focus on one of the decaying phases presented in
    Fig.~\ref{f_hautpm}. The decaying phase is characterized by a
    change of the dominant symmetry, as we can see on the
    visualizations of the magnetic field lines. The color insets
    respectively correspond to snapshots in the quadrupolar phase
    (left) and dipolar phase (right).}\label{f_ebs}
\end{figure}

Figure~\ref{f_hautpm} presents the results we obtain at
$\prandtl=2\,$.  Close to the threshold, the magnetic field still
exhibits intermittency, but the nature of the process has
significantly changed. There is now a clear distinction between two
different regimes: phases of dynamo activity separated by phases of
pure exponential decay. Both seem to alternate randomly.  When the
dynamo is active, the magnetic field still displays a quadrupolar
symmetry. In contrast, we observe the emergence of an axial dipole
during decaying phases. The change of the global symmetry of the field
coincides with the change of slope in the decaying phases [see
Fig.~\ref{f_hautpm} (bottom) and Fig.~\ref{f_ebs}]. This change of
slope is associated with a slower decay of the dipolar component over
the quadrupolar mode.
    
\section{Discussion}\label{s_dis}

\subsection{Canonical model for on-off intermittency} 

The simplest model that exhibits on-off intermittency
is~\cite{petrelis1}
\begin{equation}\label{eq_onoff}
  \dot{X} = \left[ \dist + \zeta(t)\right] X - X^3 \,,
\end{equation}
where \dist{} is the distance to the threshold, and $\zeta$ a Gaussian
white noise of zero mean value and amplitude~$D$, defined as
$\langle\zeta(t)\zeta(t')\rangle = D \delta(t-t’)$ where
$\langle\rangle$ indicates the average over realizations (ensemble
average).  In the absence of noise, the system undergoes a
supercritical pitchfork bifurcation at $\dist=0\,$.  If \dist{} is
sufficiently small, the fluctuations lead to on-off intermittency,
with bursts ($\dist +\zeta > 0$) followed by decays ($\dist +\zeta <
0$). During the off phases, one can neglect nonlinearities and write
$\dot{Y} = \dist + \zeta(t)$, with $Y=\ln(X)$.  Thus, $\ln(X)$ should
follow a random walk, with a small positive bias. Since solutions of
Eq.~\eqref{eq_onoff} mimic solutions of the magnetohydrodynamics
equations we observe in Fig.~\ref{f_onoff}, we further investigate
some properties of the model. (i) Equation~\eqref{eq_onoff} leads to a
stationary probability density function (PDF) of the
form~\cite{stratonovitch}
\begin{equation}\label{eq_pow}
  P(X) \propto X^{(2\dist/D)-1} e^{-X^2/D} \,,
\end{equation}
which diverges at the origin for $0\leqslant s=(2\dist/D)-1
<1\,$. (ii) In addition, all the moments of $X$ must follow a linear
scaling with \dist{}.  (iii) Finally, another characteristic of this
model is that the distribution of the duration of the off
phases~\toff{} follows a power law behavior, $P(\toff) \sim
\toff^{-\alpha}$, with $\alpha=-3/2$.  To compare these predictions to
our results, we rely as in~\cite{alex} on the magnetic energy density
as a global measure of the magnetic field strength.

\subsection{Predictions and results} 

\begin{figure}[htbp]
    \centering
    \includegraphics[width=0.4\textwidth,clip=true,
        trim=0 180 0 180]{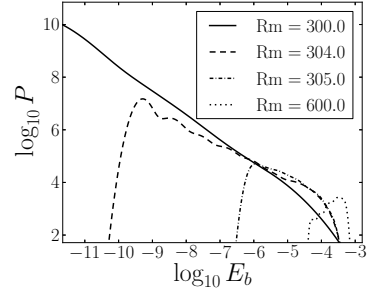}
  \caption{Estimates of the probability density functions. Statistics
    are done from time series of the magnetic energy obtained with
    ferromagnetic boundary conditions, for
    $\prandtl=0.2$.}\label{f_pdf1}
\end{figure}

\begin{figure}[htbp]
    \centering
    \includegraphics[width=0.4\textwidth,clip=true,
      trim=0 180 0 180]{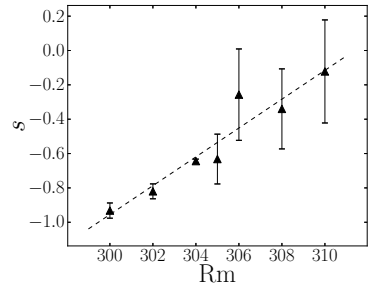}
  \caption{Fit of the coefficient $s=(2\dist/D)-1$, taking into
    account the linear domain of the PDFs in the intermittent regime
    only. Statistics are done from time series of the magnetic energy
    obtained with ferromagnetic boundary conditions, for
    $\prandtl=0.2$.}\label{f_pdf2}
\end{figure}

\begin{figure}[htbp]
    \centering
    \includegraphics[width=0.4\textwidth,clip=true,
        trim=0 180 0 180]{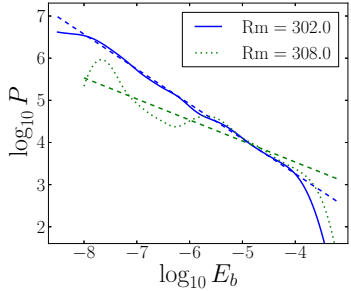}
  \caption{(Color online) Examples of the fit (dashed lines) of the
    probability density functions (solid and dotted lines).
    Statistics are done from time series of the magnetic energy
    obtained with ferromagnetic boundary conditions, for
    $\prandtl=0.2$.}\label{f_testfit}
\end{figure}

Figure~\ref{f_pdf1} shows the PDFs of the magnetic energy for a set of
simulations at different Reynolds numbers. 
At low \reymag{}, the PDF is characterized by a linear scaling on a
log-log plot. The cutoff at low energies is not predicted by the
theory, which considers the limit \mbox{$\Eb\to0\,$}.  For
$\reymag>310\,$, the magnetic energy fluctuates around a mean value
and the PDF no longer scales as a power law.  We see in
Fig.~\ref{f_pdf2} that the coefficient $s$ is proportional to the
distance to the threshold.
Examples of the fit of the exponent $s$ are
presented in Fig.~\ref{f_testfit}. 
The values of the coefficient are mainly affected by the range over
which the data are fitted. Thus we select a range as large as
possible. We then randomly sample this range with half-size
sub-intervals.  We then compute the mean slope and its standard
deviation (represented in Fig.~\ref{f_pdf2} with error bars).

\begin{figure}[htbp]
  \begin{center}
    \includegraphics[width=0.4\textwidth, clip=true, trim=0 180 0
      180]{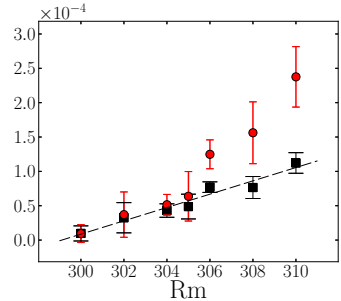}
    \caption{(Color online) Moments of the magnetic energy as function
      of the magnetic Reynolds number. Black squares represent the
      mean. The dashed line fits these data points with an error about
      10\% on the slope coefficient. The second moments (red circles)
      have been rescaled to match with the mean at
      $\reymag=300\,$.}\label{f_mom}
  \end{center}
\end{figure}

We then investigate the linearity of the moments. Figure~\ref{f_mom}
shows our results for the first and second moments of the magnetic
energy.  We see that the mean magnetic energy grows linearly as a
function of the magnetic Reynolds number.  The second moments seem to
follow the same linear trend, but only at the lower values of the
magnetic Reynolds number.  Deviations at larger values of \reymag{}
are expected, as this description is only valid in the limit $\reymag
\to \rmc\,$. The duration~$t_I$ of the time series used to compute
these values ranges from $3.2 \,10^{3}$ to $1.4 \, 10^{4}$
U.T. (values are presented in Table~\ref{t_dt}).  These integration
times are quite significant for a fully three dimensional set of
partial differential equations but are necessarily short compared to
the ones usually used with simplified models such
as Eq.~\eqref{eq_onoff}.  To quantify the uncertainty associated with
the moment values, we sampled the integration time with sub-intervals
randomly chosen. We then computed the moments on the full interval
(symbols in Fig.~\ref{f_mom}) and the standard deviation on the
sub-intervals (reported as error bars).  The sub-intervals can be set
from $t_I/4$ to $t_I/10$ without affecting these estimates.
\begin{table}[htbp]
  \begin{center}
    \begin{tabular}{c c c}
      \hline\hline
               \reymag{} &  $t_I$ & $t_I / \reymag$\\
      \hline
      300   & $1.40 \, 10^{4}$ & 46.7  \\ 
      302   & $9.06 \, 10^{3}$ & 30.0  \\ 
      304   & $9.05 \, 10^{3}$ & 29.8  \\ 
      305   & $3.33 \, 10^{3}$ & 10.9  \\ 
      306   & $7.58 \, 10^{3}$ & 24.8  \\ 
      308   & $3.20 \, 10^{3}$ & 10.4  \\ 
      310   & $5.46 \, 10^{3}$ & 17.6  \\ 
      \hline\hline
    \end{tabular}
    \caption{Duration of the time series used to compute the moments
      of the magnetic energy. The integration time $t_I$ is presented 
      in units of $\left(\chi \Delta \Omega \right)^{-1}$ 
      ($t_I$) and 
      in units of the magnetic diffusion time $\ro^2 / \eta$
      ($t_I / \reymag$). }\label{t_dt} 
  \end{center}
\end{table}

\begin{figure}[htbp]
    \centering 
    (a)\includegraphics[width=0.4\textwidth, clip=true,
      trim=0 180 0 180]{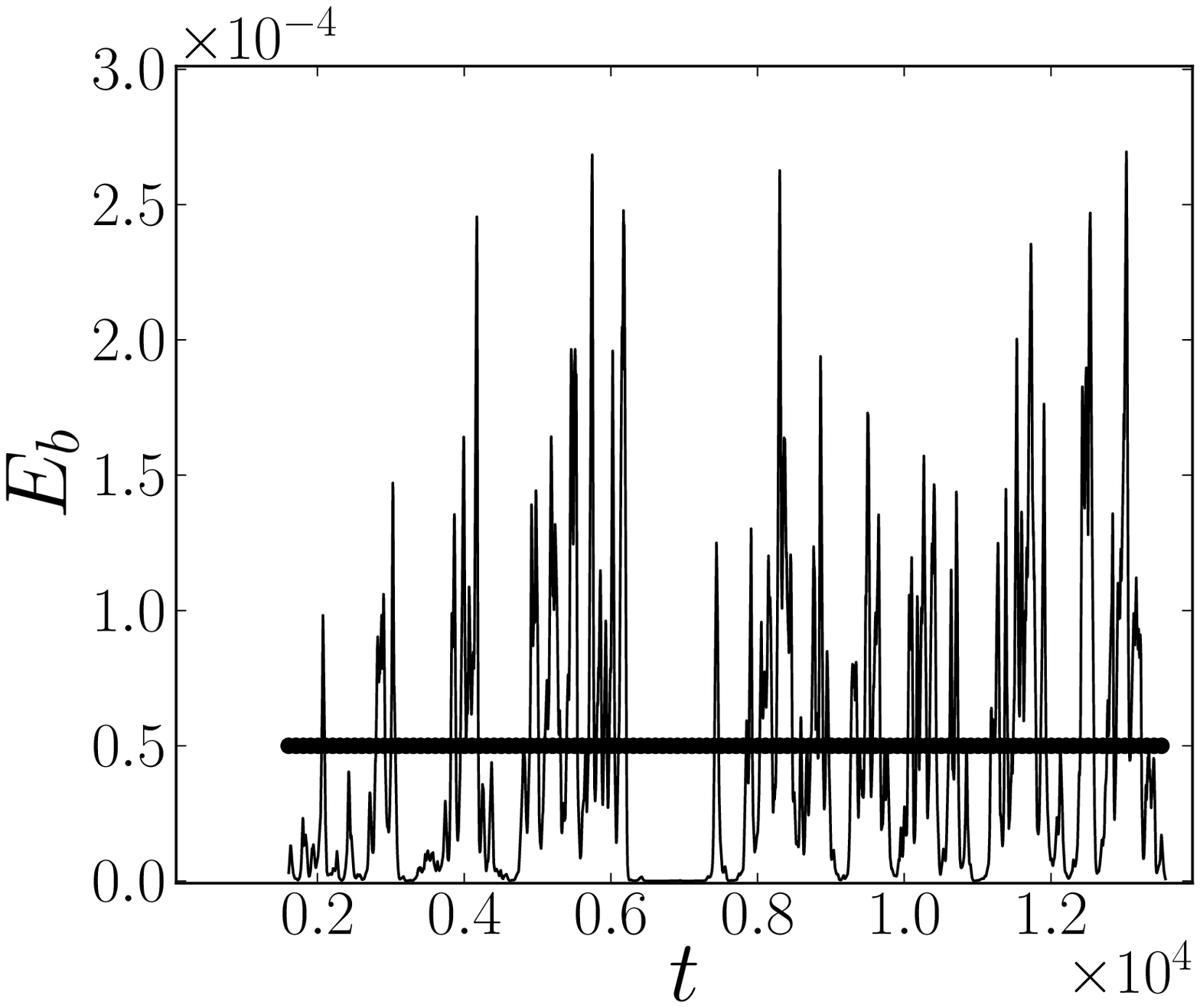}
    (b)\includegraphics[width=0.4\textwidth, clip=true, trim=0 180 0
      180]{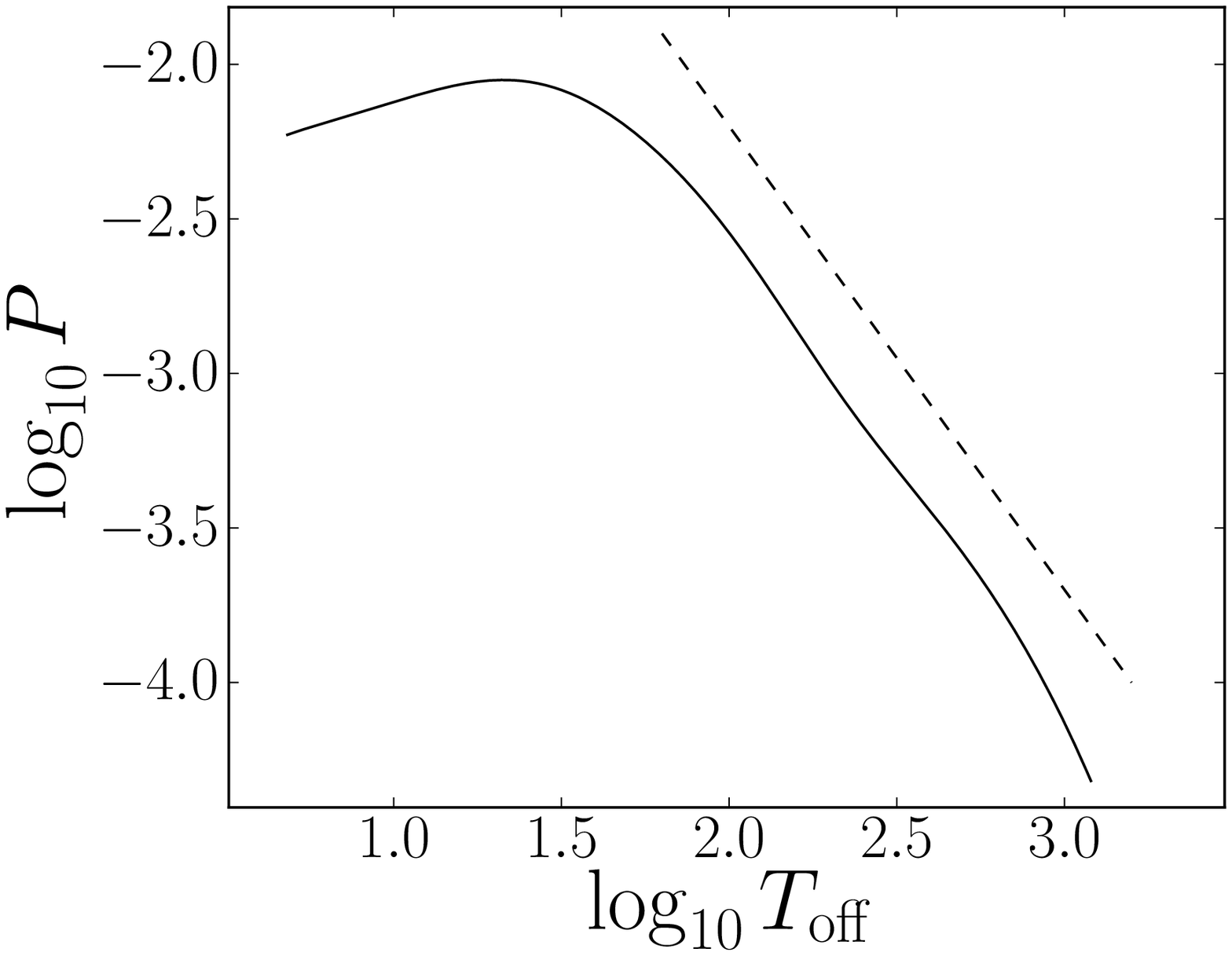}
  \caption{ Distribution of the duration of the off phases for a
    simulation performed at $\prandtl=0.2$ and $\reynolds=1470$, with
    conducting boundary conditions. We define an off phase by a
    magnetic energy below a threshold value (indicated here by the
    horizontal black line). The PDF (b) decay is close to a power law
    with the expected $-3/2$ value for the exponent (dashed
    line).}\label{f_toff}
\end{figure}

Finally, we also tested the distribution of the duration time of the
off phases. A definitive validation would require longer simulations,
in order to have a significant number of off phases. For this reason,
we can not rely on the simulations immediately above the
threshold. Despite these short-comings, an illustrative case is
presented in Fig.~\ref{f_toff}. Numerical values are given in
Table~\ref{t_cutoff}.
\begin{table}[htbp]
  \begin{center}
    \begin{tabular}{c c c c}
      \hline\hline
      \diagbox{Threshold}{Range} \ \ & [1.7 ; 3]  & [2.0 ; 3]  & [2.1 ; 3]  \\ 
      \hline
      $1.1 \, 10^{-4}$ & -1.30 & -1.48 & -1.51 \\ 
      $1.0 \, 10^{-4}$ & -1.35 & -1.48 & -1.50 \\
      $7.5 \, 10^{-5}$ & -1.40 & -1.41 & -1.40 \\
      $5.0 \, 10^{-5}$ & -1.48 & -1.52 & -1.51 \\
      $3.5 \, 10^{-5}$ & -1.51 & -1.60 & -1.65 \\ 
      \hline\hline
    \end{tabular}
    \caption{Estimate of the exponent $\alpha$ of the PDF of the
      duration of the off phases for different threshold values and
      different ranges over which the fit is done. The standard error
      on the estimate of $\alpha$ is about $1\%$. Range values
      correspond to $\log_{10} \toff$ [x axis in
        Fig.~\ref{f_toff}(b)].}\label{t_cutoff}
  \end{center}
\end{table}

To conclude, we emphasize that the predictions of the model are
consistent with the three-dimensional simulations, and thus confirm
the on-off hypothesis for the observed intermittency at low magnetic
Prandtl number.

\subsection{Simulations at higher magnetic Prandtl number}

\begin{figure}[htbp]
  \begin{center}
    \includegraphics[width=0.4\textwidth, clip=true, trim=0
      180 0 180]{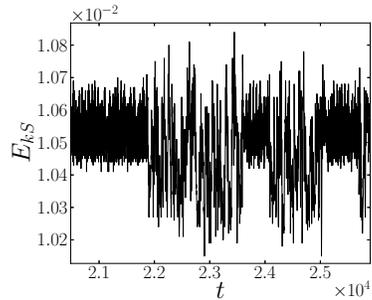}
  \caption{Time evolution of the symmetric part of the kinetic energy
    $E_{kS}$ at $\reynolds=1200$ in a purely hydrodynamic simulation.
    Intermittent transitions between ``laminar'' and more turbulent
    phases are clearly visible.}\label{f_ek}
  \end{center}
\end{figure}

The simulations we performed at $\prandtl=2$ exhibit a peculiar
behavior of the magnetic field. This can be better understood by
examining the dynamics of the flow. Indeed, we also carried out purely
hydrodynamic simulations at $\reynolds=1200$ and observed intermittent
transitions between two states. This kind of intermittent behavior of
the flow was not reported in~\cite{celine}, but has been observed
experimentally~\cite{lathrop}.
One state is characterized by larger
fluctuations of the energy as we can see in Fig.~\ref{f_ek}.
In addition, the analysis of the energy spectra reveals that the $m=3$
modes dominate over the $m=2$ modes during the ``laminar'' phases,
whereas both are of the same order during the ``turbulent'' phases.
Duration of the ``turbulent'' phases tends to increase gradually with
the increase of the Reynolds number, so that the intermittent behavior
of the flow eventually disappears and is thus no longer present in the
simulations at higher Reynolds number in which we have identified
on-off intermittency.

\begin{figure}[htbp]
  \begin{center}
    \includegraphics[width=0.4\textwidth, clip=true, trim=0
      180 0 180]{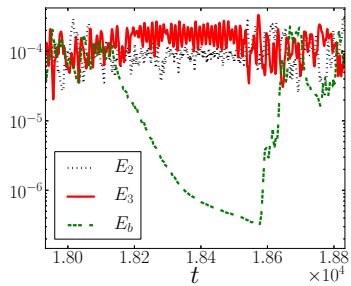}
  \caption{(Color online) Time evolution at $\reynolds=1200$ and
    $\prandtl=2$ of the total magnetic energy $E_b$ (dashed green
    line) and the kinetic energies $E_2$ (dotted black line) and
    $E_3$ (solid red line) for the $m=2$ and $m=3$ modes,
    respectively. When the latter becomes larger than the former, the
    dynamo is no longer sustained and the magnetic energy
    exponentially decays.}\label{f_em2}
  \end{center}
\end{figure}
Dynamo action is inhibited during the ``laminar'' phases (when the
$m=3$ modes dominate), which highlights the mechanism which leads to
the peculiar behavior of the magnetic field, as we can see in
Fig.~\ref{f_em2}.   In contrast, ``turbulent'' phases favor dynamo
action, and one must wait a change in the flow to see the restart of
dynamo action after a phase of decay. Moreover, in a full
magnetohydrodynamics simulation, we can artificially suppress the
$m=3$ modes of the velocity field by setting them equal to zero at
each time step. We check that it is sufficient to suppress
intermittency of the flow. Then, we observe that the phases of
exponential decay are also suppressed and the dynamo is no longer
intermittent.

\section{Conclusion}

Despite the fact that on-off intermittency has so far never been
observed in dynamo experiments, we showed that the phenomenon can
appear in numerical simulations of dynamo action using realistic
boundary conditions.  We identified in several cases the predicted
behavior of the PDF of the magnetic energy, linear scaling of the
moments, and distribution of the duration of the off phases.  In
addition, we tested these properties for three different boundary
conditions (conducting inner core with insulating outer sphere,
insulating or ferromagnetic spheres).  Finally, we pointed out a
different kind of intermittency due to hydrodynamic transitions that
appears at lower Reynolds numbers.

To explain the absence of on-off intermittency in the experiments,
several reasons have already been invoked~\cite{petrelisGAFD2007}. One
explanation could be the imperfectness of the bifurcation (due for
instance to Earth's ambient magnetic field). Since it has been shown
that low-frequency noise controls on-off
intermittency~\cite{petrelis1}, another possible explanation could be
that the low-frequency fluctuations are too small.  However, the lack
of experimental observations of on-off intermittency remains an open
question and needs further investigations.

\acknowledgments{The numerical simulations have been carried out at
  CEMAG, CINES, and MESOPSL.  We thank S.~Fauve, F.~Pétrélis, and
  M.~Schrinner for fruitful discussions and comments. We are most
  grateful to C.~Gissinger for technical assistance.}

\clearpage
\bibliography{onoff,vks,num}

\end{document}